\documentclass[
 reprint,
 amsmath,amssymb,
 aps,
 floatfix,
]{revtex4-2}

\usepackage{graphicx}
\usepackage{tabularx}
\usepackage{dcolumn}
\usepackage{bm}
\usepackage[usenames,dvipsnames]{color}

\begin{document}

\preprint{APS/123-QED}

\title{ Electrostatic modulation of excitonic fluid in GaN/AlGaN quantum wells by deposition of few-layered graphene and nickel/gold films. }

\author{R.~Aristegui, P.~Lefebvre, C.~Brimont, T. Guillet, M. Vladimirova}
\affiliation{Laboratoire Charles Coulomb (L2C), University of Montpellier, CNRS, Montpellier, France}

\author{I. Paradisanos, C. Robert, X. Marie, B. Urbaszek}
\affiliation{LPCNO, Universit\'e de Toulouse, INSA-CNRS-UPS, Toulouse, France}

\author{S. Chenot, Y. Cordier, B. Damilano}
\affiliation{{CRHEA, Universit\'e C\^ote d’Azur, CNRS, Valbonne, France}
}

\begin{abstract}

Excitons hosted by GaN/(Al,Ga)N quantum wells (QWs) are spatially indirect due to the giant built-in electric field that separates electrons and holes along the growth direction. 
This electric field, and thus exciton energy, can be reduced by depositing metallic layers on the sample surface. Using spatially-resolved micro-photoluminescence spectroscopy we compare the effects of two different materials, Nickel/Gold (NiAu) and few-layered graphene (FLG), on the potential landscape experienced by the excitons. We are able to (i) determine the potential barriers imposed on QW excitons by deposition of FLG and NiAu to be $14$ and $82$~meV, respectively, and (ii) to evidence their impact on the exciton transport at appropriate densities. 
Optical losses and inhomogeneous broadening induced by deposition of NiAu and FLG layers are similar, and their joined implementation  constitute a promising tool for electrostatic modulation of IX densities even in the absence of any applied electric bias.


\end{abstract}

\maketitle

\section{Introduction}
\label{sec:Introduction}

Indirect excitons (IXs) are bosonic quasi-particles composed of an electron and a hole confined in two spatially separated planes, but still bound by Coulomb attraction ~\cite{Lozovik1976}. 
They are generated optically either in specially designed, epitaxially grown semiconductor heterostructures ~\cite{Chen1987} or,  more recently, in van der Waals assemblies of transition metal dichalcogenides (TMDs)~\cite{Jiang2021,Gerber2019,Rivera2015}.
In heterostructures  where IX is not the lowest energy excited state, observation of its optical emission requires application of an external electric bias. 
 {IX} lifetime can be as long as several microseconds, so that they can thermalize with the crystal lattice and reach the temperature of quantum degeneracy prior to recombination ~\cite{MazuzHarpaz2017,Butov2016,Wang2019}. Additionally, IXs possess an substantial (as compared to dipolar molecules) dipole moment $D=d\cdot e$, where $e$ is electron charge, and $d$ is electron-hole separation along the structure axis. This enables their in-situ manipulation through electrostatic gates ~\cite{Unuchek2018,Jauregui2019,High2007}.
 A rich variety of potential landscapes for GaAs-hosted IXs have been engineered by depositing metallic electrodes on the sample surface~\cite{High2009,Remeika2015,Butov2017,Ciarrocchi2022}. This has led to the observation of such remarkable phenomena as quantum coherence, darkening, superfluidity, supersolid ordering and Mott insulating states~\cite{High2012,Shilo2013,Cohen2016,Anankine2017,Misra2018,Lagoin2021,Lagoin2022}, as well as to the realisation of original excitonic devices \cite{High2007,Baldo2009,Grosso2009,Andreakou2014,Butov2017}. 

\begin{figure*}
	\includegraphics[width=6in]{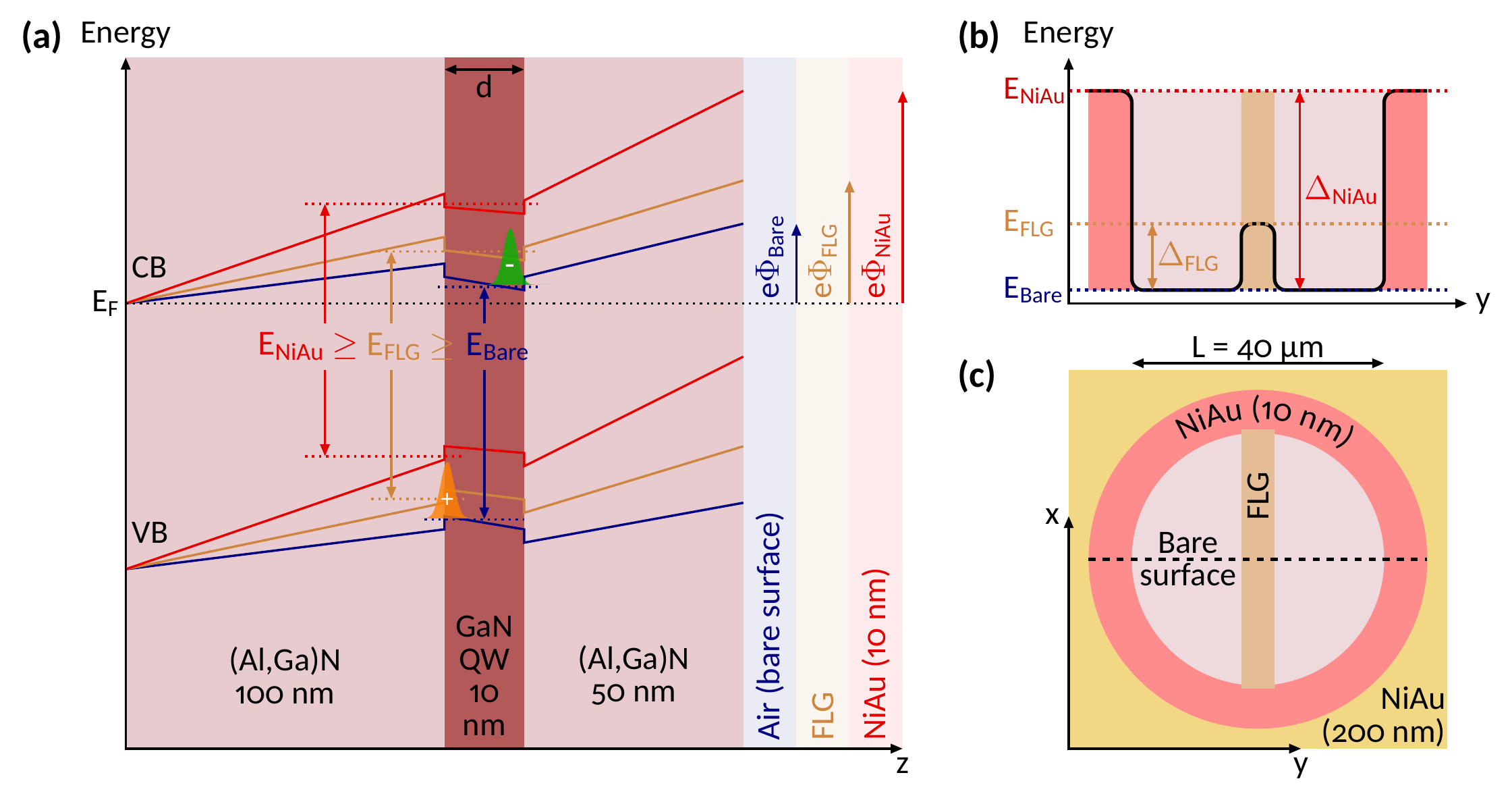}
	\caption{(a) Sketch of the GaN/(Al,Ga)N QW band diagram in the absence (blue lines) and in the presence of either FLG (brown lines) or NiAu (red lines). Corresponding surface potential energies, energy levels, and interband transition energies are shown with the same color code. Spatially separated electron and hole wavefunctions within exciton are shown schematically. Black dotted line indicates the Fermi energy. (b) In-plane potential experienced by IXs propagating along the dashed line in the environment patterned with FLG and NiAu shown in (c). Vertical arrows indicate the potential energy barriers experienced by IXs under FLG (brown) and NiAu (red) layers. (c) Top-view of the sample surface with thick NiAu, thin NiAu, and FLG layers deposited on the heterostructure surface. Horizontal arrow shows the diameter of the bare surface area.
	}
		\label{fig:intro}
\end{figure*}
 

Excitons hosted by wide polar GaN/(Al,Ga)N quantum wells (QWs) offer a  versatile platform with original features for manipulating IX dynamics
\cite{Fedichkin2015,Fedichkin2016}.
Indeed, in these QWs an electric field reaching hundreds of kV/cm along the growth direction builds up due to the difference in spontaneous and piezoelectric polarization between the well and the barrier ~\cite{Leroux1998,Grandjean1999}. This field alters the band structure dramatically, see Fig. \ref{fig:intro}~(a), separating in the growth direction the electron and the hole that constitute an exciton. As a consequence, excitons are spatially indirect even in the absence of any external bias ~\cite{Fedichkin2015,Fedichkin2016}. This make it possible to avoid spurious currents induced by an applied electric bias that are known to be detrimental to the IX coherence ~\cite{Honold1989,Koch1995,Anankine2018}. As a different approach, as we have shown previously, IX potential landscape can be patterned in these QWs by simple deposition of Nickel/Gold (NiAu) layers ~\cite{Chiaruttini2019,Aristegui2022}. This is illustrated in Fig. \ref{fig:intro}. NiAu layer raises the surface potential $\Phi$ and thus reduces the built-in electric field $F_z$ across the device \cite{Chiaruttini2019,Rickert2002,Miura2004,Schmitz1996}. Therefore, the IX emission energy in the presence of the NiAu electrode (red arrow) increases with respect to the bare surface (blue arrow). Remarkably, unlike in traditional GaAs-based systems, here IXs accumulate in the QW regions below the bare surface, which are not covered by the metal, see Fig. \ref{fig:intro}~(b-c). This ensures better photon collection efficiency. While an external bias can still be applied to the electrodes in order to manipulate the electrostatic environment~\cite{Aristegui2022}, it is appealing to find zero-bias approaches to enrich the available potential landscapes for IXs, e.g. by changing the composition of the electrodes.

In this paper, we compare few-layered graphene (FLG) and NiAu layers deposited on the surface of GaN/(Al,Ga)N QW in terms of their ability to modify IX potential landscape. We demonstrate that a combination of these to materials generates, without any applied bias, a trapping potential with two different barrier heights.


We employ low-temperature spatially-resolved micro-photoluminescence ($\mu$PL) spectroscopy to investigate in detail the IX emission and transport in three different regions of the sample: from the bare surface and from the regions covered by FLG and NiAu. Careful analysis of the $\mu$PL allows us to determine the IX energy near the zero-density limit in each of the three regions, and to map the potential landscape experienced by IXs. The outcome is schematized in Fig.~\ref{fig:intro}~(b-c): the potential profile in Fig.~\ref{fig:intro}~(b) corresponds to the dashed line in the surface pattern sketched in (c). {It appears that FLG deposition results in potential barriers $\Delta_{\mathrm{FLG}}$ that are almost $6$ times smaller as compared to NiAu electrodes, $\Delta_{\mathrm{NiAu}}$. Nevertheless, they are high enough to affect IX transport along the QW plane}. We also investigate optical losses and emission broadening induced by the deposition of FLG and NiAu layers, which are found to be very similar.

\section{Sample and experimental setup}
\label{sec:samples}

The studied sample is a GaN QW of $d=10$~nm thickness. It is sandwiched between two Al$_{0.08}$Ga$_{0.92}$N barriers, $50$- (top) and $100$~nm-thick (bottom), as shown in Fig.~\ref{fig:intro}~(a). It is grown by molecular beam epitaxy on a ($0001$)-oriented GaN free-standing substrate (LUMILOG) with $\sim 2 \times 10^{7}$~cm$^{-2}$ threading dislocation density, followed by a $600$~nm-thick GaN buffer layer.

The pattern deposited on the sample surface by optical lithography is sketched in Fig.~\ref{fig:intro}~(c). It consists of the $220$~nm-thick  NiAu layer ($20$~nm-thick nickel and $200$~nm-thick gold) on top of the $10$~nm-thick NiAu layer ($5$~nm-thick nickel and $5$~nm-thick gold) with circular apertures. The inner diameter of the circular aperture in the thinner NiAu layer is $L = 40~ \mu$m. 

Subsequently, a viscoelastic deterministic transfer process of exfoliated highly-ordered pyrolytic graphite on a polydimethylsiloxane (PDMS) stamp is employed to deposit FLG across the lithographic pattern, making contact  with the bare surface and the NiAu electrode \cite{Castellanos-gomez2014}.
It is important to point out that the FLG deposition process is considerably simpler than patterning of the samples with electrodes by optical lithography, making this a practical and appealing approach.



The sample is housed in a cryostat and cooled down {to $10$~K}. Excitons are generated by a continuous-wave laser at $3.49$~eV. This energy is slightly above the  bandgap of GaN, but well below Al$_{0.08}$Ga$_{0.92}$N barriers gap. The laser beam is tightly focused on the sample surface. The $\approx1$~ $\mu$m-diameter spot can be positioned in different areas of the sample plane, {\it e.g.} on the semi-transparent NiAu layer  or on the bare surface. 

The PL from the sample surface is collected using a microscope objective. The resulting PL image across $250\times 1$~$\mu$m sample surface is analysed by a spectrometer equipped  with a $1200$-gr/mm grating centred at $3.26$~eV. A $2048 \times 512$ pixel CCD camera captures the direct spatial and spectral PL image with spatial and spectral resolution of $\approx 1$~$\mu$m and $\approx 1$~meV, respectively.

\begin{figure}
	\includegraphics[width=3.2in]{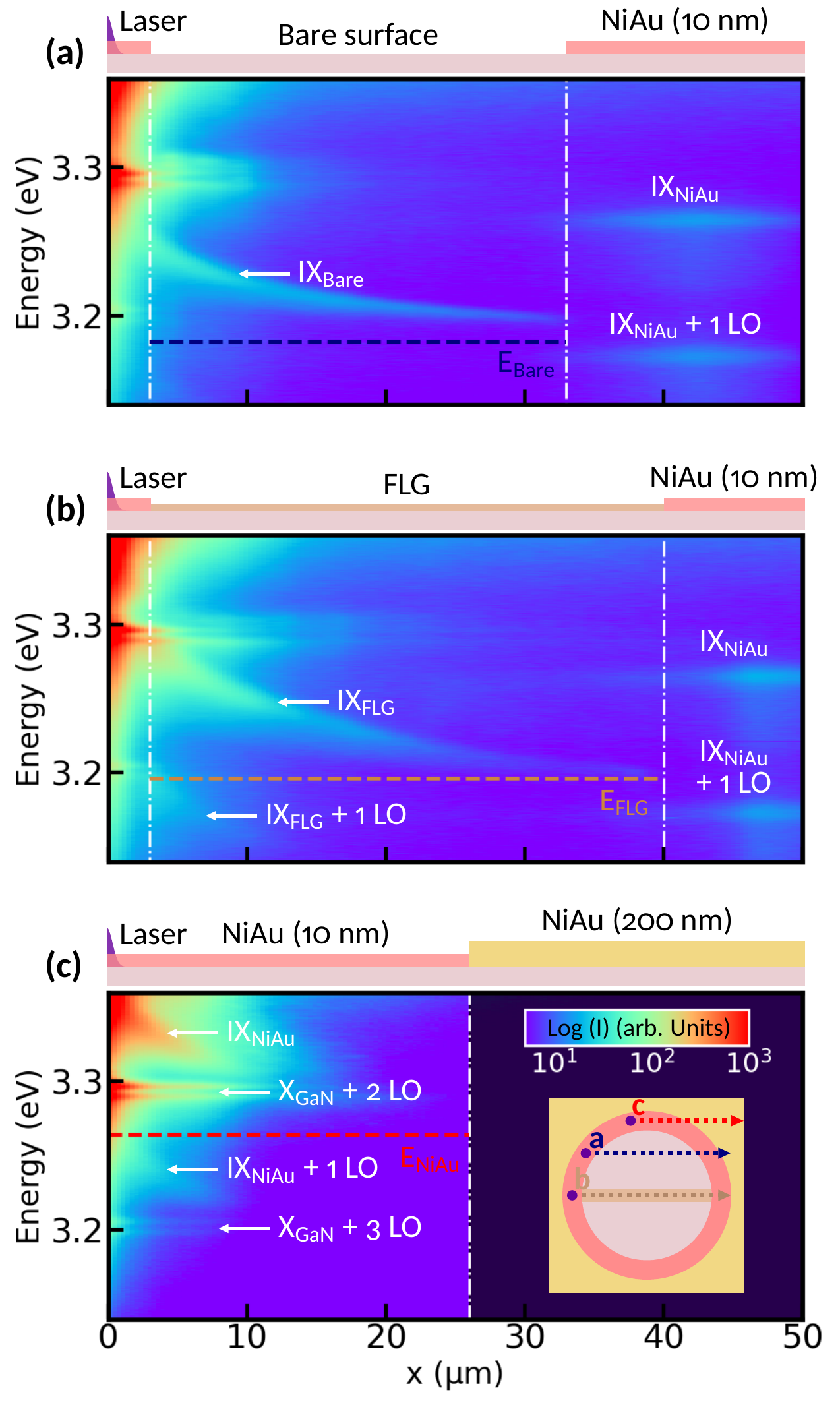}
	\caption{ PL spectra (color-encoded in log scale) measured along three different lines across the surface as shown in the inset. Excitation power densities are $P=110$~kW/cm$^2$ (a, b) and $P=140$~kW/cm$^2$ (c). The corresponding surface state and the laser position are sketched above each spectral map. The energy values at the zero-density limit for IX under the bare surface (blue), FLG (brown), and NiAu (red) are indicated by the dashed lines. 
	}
		\label{fig:SpectralMaps}
\end{figure}

It is important to note that spatial resolution is essential for this work. Indeed, dipole-dipole repulsion between IXs pushes them away from the excitation spot. It has been shown that IX lifetime can be as long as microseconds \cite{Leroux1998,Grandjean1999,Lefebvre2004,Fedichkin2015}. As a result, under steady state excitation IX emission spreads over tens or even hundreds of microns away from the excitation spot. The spatial extension of the IX emission depends on the laser power and the surface potential that are altered by the deposition of FLG and NiAu, as the experiments described below demonstrate \cite{Fedichkin2016,Chiaruttini2019}. Moreover,  IX emission energy and intensity depend strongly on the IX density, due to self-induced screening of the built-in electric field \cite{Lefebvre2004}. Therefore, density correlations within the excitonic fluid must be taken into account to unravel the population-induced and surface deposition-induced effects\cite{Lefebvre2004,Chiaruttini2019}. The relations between surface potential $\Phi$, exciton density $n$, and emission properties are explored in the next section. 


\section{Experimental results and discussion}
\label{sec:experiments}

\subsection{Patterning IX potential with FLG and NiAu}
\label{subsec:PotentialBarriers}

The aim of this section is to determine and compare the height of the potential barriers experienced by IXs following the deposition of FLG ($\Delta_{\mathrm{FLG}}$) and NiAu ($\Delta_{\mathrm{NiAu}}$) layers on the sample surface. 
Because the density-dependent blueshift can be larger than this potential barrier, we cannot simply rely on the comparison between  the spectra recorded under the excitation spot focused  on the bare surface and on each of the  investigated layers. As much more reliable determination of the potential barrier, we propose to analyse the entire spatial and spectral map of the IX emission upon excitation with variable power density in different areas of the sample. 

The experiments are conducted in three configurations sketched in the inset of Fig.~\ref{fig:SpectralMaps}~(c). They allow us to analyse the emission from the bare (Al,Ga)N surface (a), the surface covered by FLG (b), and the surface covered by NiAu (c). 
The laser is always positioned on the NiAu semi-transparent electrode to minimise laser-induced heating of IXs in configurations (a) and (b). To exclude heating-induced artefacts in the configuration (c) we restrict the analysis of the IX emission to the spatial areas situated more than $5~\mu$m away from the laser spot position.

Fig.~\ref{fig:SpectralMaps}~(a-c) shows color-encoded PL spectra, measured at various distances from the excitation spot in these three configurations.
The laser spot is positioned at $x=0$. At this point in all configurations we detect spectrally broad emission from GaN buffer, on top of the strongly broadened IX emission as well as its longitudinal optical (LO) phonon  replica (IX+1LO), and two double peaks corresponding to two and three  LO phonon replica of free and donor-bound excitons in GaN buffer ($E_{\mathrm{LO}}\approx 90$~meV in GaN) \cite{Dingle1971,Kornitzer1999,Reshchikov2021}.
%
Light at LO replica energies is observed at distances of up to $10$~$\mu$m distances from the excitation. We speculate that these photons represent a fraction of the high intensity emission under the excitation spot, that is scattered by the surface imperfections.

Our main interest here is the excitonic emission at distances $x \ge 5$~$\mu$m. 
It can be easily identified because its energy decreases with the distance away from the excitation spot.
The reduction in the energy indicates that the  density of IXs decreases due to radial dilution and optical recombination. 
This kind of emission pattern is typically observed in spatially resolved PL experiments in various IX-hosting systems \cite{Kuznetsova2015,Fedichkin2015,Dorow2018}.
We can also distinguish the first LO phonon replicas of IX that follows the same emission pattern but $90$~meV below IX emission.

The emission observed at $3.264$~eV and at $3.174$~eV in Fig.~\ref{fig:SpectralMaps}~(a) and (b) at $x>33$~$\mu$m and at $x>40$~$\mu$m, respectively, deserves a special comment. 
The corresponding areas are covered by NiAu and situated at significant distance from the excitation spot. 
%
It will be shown below that, because the built-in electric field $F_z$ is much weaker under NiAu than in the bare surface areas, $F_{\mathrm{NiAu}} \ll F_{\mathrm{Bare}}$, the corresponding electron-hole wavefunctions overlap is exponentially higher, leading to a much brighter emission even at very low IX densities.
Thus, we  anticipate the identification of these lines as the zero-density excitonic transition in NiAu-covered region and its $1$~LO phonon replica.

\begin{figure}
	\includegraphics[width=3.4in]{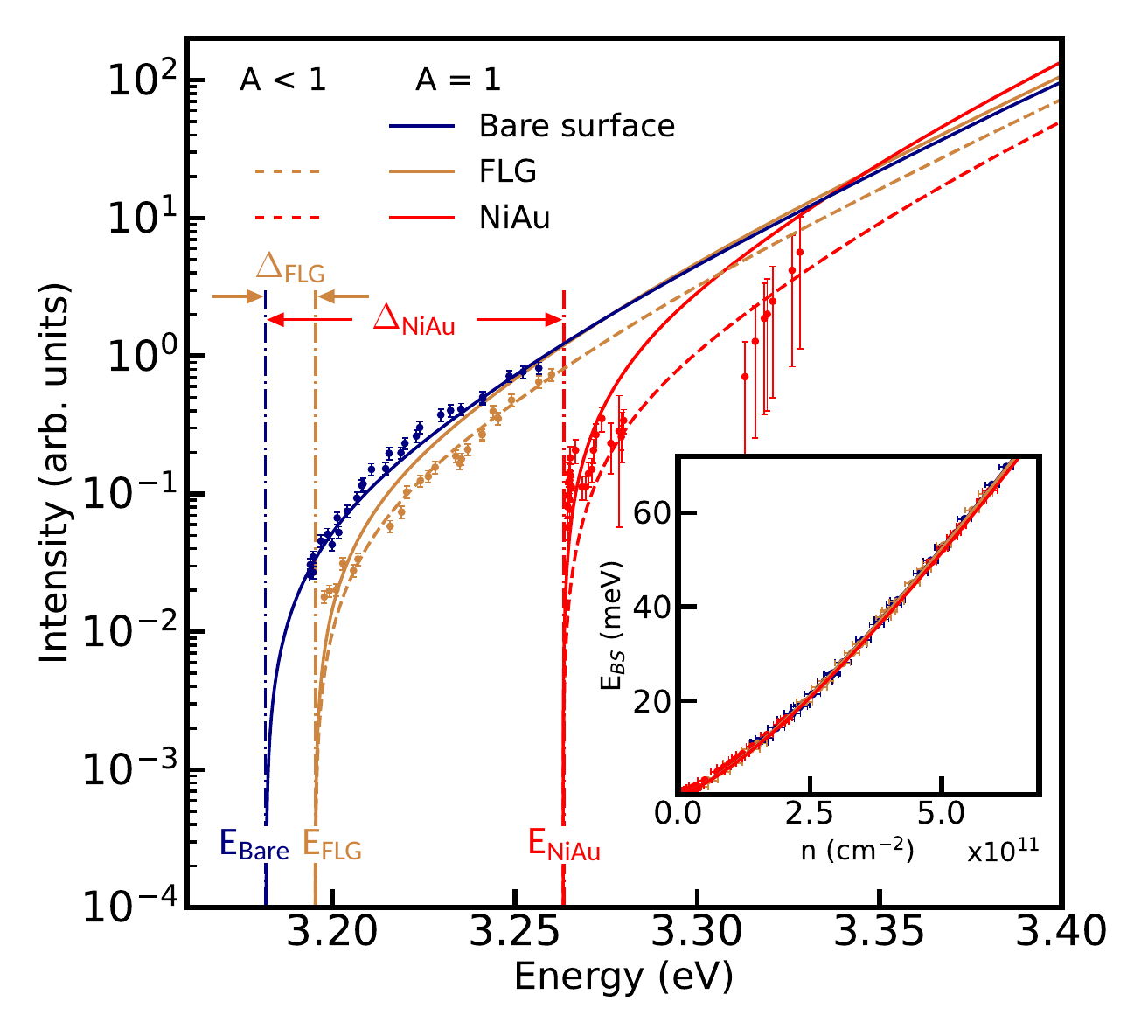}
	\caption{Spectrally integrated intensity of the IX emission (symbols) as a function of its peak energy, extracted from spatially-resolved PL measurements (like those in Fig.~\ref{fig:SpectralMaps}). Different points correspond to different positions in the plane of the QW and different excitation powers. Measurements under bare surface, FLG and NiAu are shown in blue, brown, and red, respectively. Dashed lines are the results of the fitting to the data using Eqs.~\ref{eq:E_IX}-\ref{eq:I_n} with the parameters given in Table \ref{tab:ParamaterFit}. Solid lines show the same model but assuming $100$~\% transmission of FLG and NiAu (A=$1$). Vertical lines point the zero-density energies extracted from the fit. Inset: the relation between IX density and the emission blueshift used in the model (same color code).
	}
		\label{fig:I_E}
\end{figure}

Fig.~\ref{fig:I_E} gathers the spectrally-integrated IX emission intensities measured across the three areas of the sample (bare surface, FLG and NiAu covered). The experiments are identical to those shown in Fig.~\ref{fig:SpectralMaps}, but various excitation powers are considered (from $P = 7$ to $160$~kW/cm$^{2}$). The intensities are reported as a function of the corresponding emission energies. Both are determined  by fitting the phenomenological function introduced in Ref.~\cite{Chiaruttini2021} to the measured PL spectra at different distances from the excitation spot.

The relation between IX emission energy and density can be understood in terms of the model presented below.
The energy $E_{\mathrm{IX}}$ of the density-dependent IX emission can be described as ~\cite{Fedichkin2015,Fedichkin2016,Lefebvre2004}:
\begin{equation} 
E_{\mathrm{IX}}(n) = E_0 + E_{\mathrm{BS}}(n),
 \label{eq:E_IX}
\end{equation}
where $E_{0}$ represents the zero-density exciton energy. It depends on the sample structure, state of the surface (bare surface, FLG or NiAu covering) and the resulting surface potential energy $e\Phi$, {\it cf} Fig.~\ref{fig:intro}~(a). By solving self-consistently Schr\"{o}dinger and Poisson equations, $E_0$ can be directly related to the surface potential $\Phi$ (which is controlled by the properties of the Schottky contacts) and the resulting electric field $F_z$ in the structure. We use NextNano++ software~\cite{Birner2007} for this calculation. 

The second term in Eq.~\ref{eq:E_IX}, $E_{\mathrm{BS}}$, is the density-dependent blueshift of the IX energy. Its density dependence is governed by the exciton built-in dipole moment $D$ and the correlations of the density within exciton fluid ~\cite{Laikhtman2009,Zimmermann2010,Ivanov2010,Dang2020,Chiaruttini2021,Lagoin2022}. It can be modelled as 
\begin{equation} 
E_{\mathrm{BS}}(n) = n \phi_0 f(n).
 \label{eq:E_BS}
\end{equation}
Here $n \phi_0$ accounts for the screening of the initial built-in electric field $F_z$ induced by the
 charge density, $n$, evenly distributed across the QW interfaces.
This density, $n$, can be deduced from the self-consistent solution of coupled Schrodinger and Poisson equations. 
The density dependent function $f(n)<1$ accounts for the density correlations within the IX fluid. It can be calculated as described in Refs.~\onlinecite{Zimmermann2010,Chiaruttini2021}.

\begin{table*}

\begin{ruledtabular}

\begin{tabular}{c|c|c|c|c|c}

Parameter & Meaning & Units & Bare surface & FLG & NiAu \\ 
\noalign{\smallskip} \hline \noalign{\smallskip}
$E_{0}$ & zero-density IX energy & eV & $3.182 \pm 0.002$ & $3.196 \pm 0.003$ & $3.264 \pm 0.005$  \\
          &     &   &       &     &        \\ 
$\phi_{0}$ &  mean-field IX & meV $ \times$ cm$^{2} \times 10^{-11}$ & $14.8 \pm 0.5$ & $14.7 \pm 0.5$ &  $13.9 \pm 0.5 $ \\ 
           &  interaction energy  &   &       &     &        \\ 
           &     &   &       &     &        \\ 
$\gamma$ &  density dependence of the  & cm$^{-2}\times 10^{11}$ & $2.44 \pm 0.1  $ & $2.45 \pm 0.1 $ & $2.46 \pm 0.1$  \\ 
          &   IX emission intensity (Eq.~\ref{eq:I_n})   &   &       &     &        \\ 
          &     &   &       &     &        \\ 
$\Omega_0$ &  squared overlap of the  & $\times 10^{-7}$ & $ 4.5 \pm 0.2$ & $7.6 \pm 0.2$  & $74 \pm 2 $ \\
          & electron and hole  wavefunctions    &   &       &     &        \\ 
          &     &   &       &     &        \\ 
$A/A_{\mathrm{Bare}}$ & optical transparency &  & $1$ & $0.68 \pm 0.1$ & $0.37 \pm 0.1$ \\
          &     &   &       &     &        \\ 
$\Phi$ & surface potential   & V & $1$ & $1.2 \pm 0.1$ & $2.38 \pm 0.1$ \\ 
          &     &   &       &     &        \\ 
$F_z $ &  built-in electric field  & kV/cm & $458$ & $446$ & $375$ \\ 
          &     &   &       &     &        \\ 

$\Gamma_i $ & inhomogeneous broadening & meV & $5.7 \pm 0.5$ & $10.5 \pm 0.5$  & $10.5 \pm 0.5$ \\ 
          &     &   &       &     &        \\ 
$\sigma $ & IX scattering crossection & meV $ \times$ cm$^{2} \times 10^{-11}$ &  $3.06 \pm 0.1 $   & $3.06 \pm 0.1 $   &  $3.06 \pm 0.1 $ \\ 
\noalign{\smallskip}
\end{tabular}

\end{ruledtabular}

\caption{Parameters extracted from the modeling of the IX emission energy, intensity, and linewidth in the  three regions of the sample plane: bare surface, FLG and NiAu.}


\label{tab:ParamaterFit}
\end{table*}

Both terms in Eq.~\ref{eq:E_IX} depend on the surface potential. Therefore, to identify precisely these two contributions and reliably determine $E_{\mathrm{0}}$ in three different areas of the excitonic landscape, we rely of the fact that under a given surface potential $\Phi$ (or, equivalently, corresponding value of the built-in electric field, $F_z$) not only the emission energy, but also the integrated intensity $I_{\mathrm{IX}}$ is density dependent: ~\cite{Fedichkin2016,Liu2016,Chiaruttini2019,Chiaruttini2021,MazuzHarpaz2017,Aristegui2022}:
\begin{equation}
I_{\mathrm{IX}}(n) = A \frac{n}{\tau_0} \mathrm{exp }\left ( \frac{n}{\gamma } \right ).
 \label{eq:I_n}
\end{equation}
Here $A$ is the proportionality factor that we choose to be equal to unity for the bare surface regions. It is introduced to account for eventual IX emission losses induced by FLG and NiAu deposition ($A_{\mathrm{NiAu, FLG}}/A_{\mathrm{Bare}}<1$). The radiative lifetime at zero density, $ \tau_0 $, is inversely proportional to the squared electron-hole overlap integral for the zero-density IX transition, $\Omega_0$. Finally, $\gamma$ characterises exponential reduction of this integral with increasing IX density. 
Both $\Omega_0$ and $\gamma$ can be deduced from the self-consistent solution of Schrodinger and Poisson equations, see Table~\ref{tab:ParamaterFit}.

Eqs.~\ref{eq:E_IX}-\ref{eq:I_n} allows us to model the dependence of the IX emission intensity on its energy. The only fitting parameters are the proportionality coefficient $A$ and the surface potential energy $e\Phi$. The latter yields the values of the built-in electric field $F_z$ and, via solution of the Schr\"{o}dinger/Poisson equations, the zero-density energy $E_0$ in each of the three areas of the quantum well: under the bare surface, in the FLG- and in NiAu-covered regions.
The fit of this model to the data is shown by solid lines in Fig.~\ref{fig:I_E}, while the corresponding density dependence of the IX emission blueshift is given in the inset. The resulting values of the fitting parameters, including the surface potential energies, as well as the calculated values of $F_z$, $\phi$, $\gamma$ and $\Omega_0$ in each of three areas are given in Table~\ref{tab:ParamaterFit}. 
Note  that the values of $e\Phi$ in different areas of the sample are consistent with (yet quite broadly dispersed) values from the literature ~\cite{Fisichella2014,Pandit2021,Giannazzo2019,Kumar2016}.

By comparing the zero-density transition energies in the three areas of the sample plane we deduce the potential barriers experienced by IXs in the areas covered by FLG and NiAu: 

\begin{eqnarray}
\Delta _{\mathrm{FLG}}=E _{\mathrm{FLG}}-E_{\mathrm{Bare}}=14 \pm 3~\mathrm{meV} \\ 
\Delta_{\mathrm{NiAu}}=E_{\mathrm{NiAu}}-E_{\mathrm{Bare}}=82 \pm 5~\mathrm{meV}
 \label{eq:Delta}
\end{eqnarray}

This means that both FLG and NiAu modify the potential landscape of IXs in the plane of the QW, although FLG-induced potential barrier is almost six times smaller. 
The result for NiAu is consistent  with our previous evaluations \cite{Chiaruttini2019,Aristegui2022}.
Because IXs screen out the built-in potential according to Eqs.~\ref{eq:E_IX}-\ref{eq:E_BS}, we can estimate the IX density such that the in-plane potential  barrier induced by deposition of FLG or NiAu is screened completely. 

This yields $n_{\mathrm{FLG}}\approx 1.7 \times 10^{11} $~cm$^{-2}$ for FLG, and $n_{\mathrm{NiAu}}\approx 7.1 \times 10^{11} $~cm$^{-2}$ for NiAu.
Assuming that excitonic Mott transition is expected to take place in the range between $2$ and $6\times 10^{11}$~cm$^{-2}$ \cite{Rossbach2014,Chiaruttini2021}, even a FLG-induced in-plane potential barrier could be sufficient to reach a significant accumulation of IXs.

\subsection{Optical transparency of FLG and NiAu}
\label{subsec:PhotonLosses}

In order to compare FLG and NiAu  in terms of {transparency} we examine the IX emission intensity shown in Fig.~\ref{fig:I_E} and the coefficient $A$ deduced from fitting Eqs.~\ref{eq:E_IX}-\ref{eq:I_n} to these data.
Note that taking into account deposition-induced changes of $E_0$ and $\Omega_0$ is essential to capture the transparency reduction correctly. 
Indeed, 
these parameters determine the density of IXs corresponding to the emission at a given energy, and thus different expected emission intensities in the areas with bare surface and in the areas covered by FLG or NiAu, see Eq.~\ref{eq:I_n}. 

The ratio $A/A_\mathrm{Bare}$ characterises IX emission losses induced by the deposition of these layers. 
Our results, see Table~\ref{tab:ParamaterFit}, indicate $\approx 70\pm 3\%$ transparency with FLG and $\approx 40\pm 3\%$ with NiAu, comparable with, but slightly below the values from the literature \cite{Jo2010,Zhang2022}.
The dashed lines in Fig.~\ref{fig:I_E} show the expected intensity in the case of $100$~\% transparency of FLG and NiAu. 
This result, however, should be exploited with care, because (i) optical transparency of FLG strongly depends on its thickness and deposition conditions, and (ii) only limited number of measurements is available in the NiAu-covered areas at high IX density.

\subsection{Spectral broadening of IX emission induced by surface patterning}
\label{subsec:InhomoBroad}

\begin{figure}%
	\includegraphics[width=3.4in]{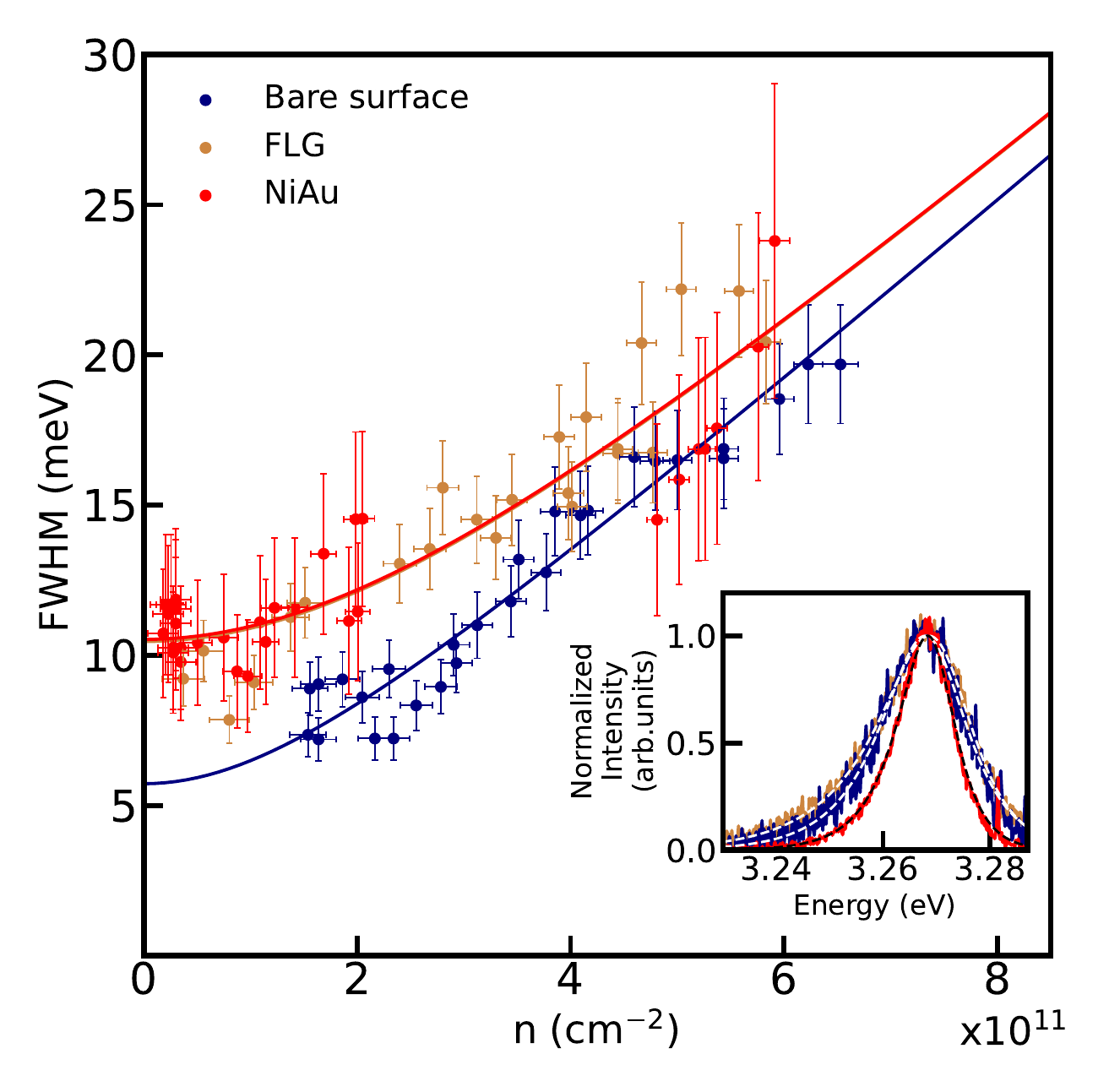}
	\caption{Density dependence of the IX emission linewidth. Data extracted from spatially-resolved PL measurements (like those if Fig.~\ref{fig:SpectralMaps}). The corresponding relation between IX density and the PL blueshift is shown in the inset of Fig.~\ref{fig:I_E}. Different points correspond to different positions in the plane of the QW and different excitation powers. Measurements under bare surface, FLG and NiAu are shown in blue, brown, and red, respectively. Lines show the fits to the data using Eq.~\ref{eq:FWHM}. Inset: IX spectra at similar emission energies measured in three different areas of the sample (same color code) and their fittings. The corresponding IX densities are  $n_{\mathrm{Bare}}\approx 7.21 \times 10^{11} $~cm$^{-2}$, $n_{\mathrm{FLG}}\approx 6.52 \times 10^{11} $~cm$^{-2}$ and $n_{\mathrm{NiAu}}\approx 0.87 \times 10^{11} $~cm$^{-2}$ for the spectrum measured under bare surface, FLG and NiAu, respectively. 
	}
		\label{fig:FWHM}
\end{figure}

Not only IX emission intensity, but also its spectral width is density dependent. 
Let us first compare three IX spectra measured at similar energies but in different areas of the sample, see Fig.~\ref{fig:FWHM},~inset. 
One can see that the  NiAu-covered region presents the narrowest emission spectrum, and the broadest one comes from under the bare surface.
This may seem surprising at first sight, but can be simply understood if we consider the density dependence of the emission energy. 
 
We represent in Fig.~\ref{fig:FWHM} the full width at half maximum (FWHM) of the IX spectra measured in the same set of experiments as in Fig.~\ref{fig:I_E}. The FWHM is shown as a function of IX density, deduced from the modeling using Eqs.~\ref{eq:E_IX}-\ref{eq:I_n}. 
One can see that (i) for a given density the FWHM is the smallest under the bare surface, and the largest under NiAu; (ii) for each of the three surface areas, FWHM increases with density. 
This latter effect can be interpreted as a homogeneous broadening induced by exciton-exciton collisions. It is expected to increase linearly with IX density \cite{Honold1989,Voros2009,Gribakin2021}:
\begin{equation}
\Gamma_h=\Gamma_{0}+\sigma n.
  \label{eq:gammahom}
\end{equation}
Here $\Gamma_{0}=\hbar/\tau_0$ is the zero density radiative emission rate, it is negligibly small in our system, $\Gamma_{0}<0.1$~meV, and $\sigma$ is the exciton-exciton scattering cross-section.
Because $\Gamma_{0}$ is very small, the linewidth at low density is dominated by the inhomogeneous broadening of the IX transition, $\Gamma_i$, mainly due to alloy composition fluctuations in the barriers.
Thus, we can describe the density dependence of the linewidth as:
\begin{equation}
\mathrm{FWHM}=\sqrt { \Gamma_h^2 +\Gamma_i^2}. 
\label{eq:FWHM}
\end{equation}

Lines in Fig.~\ref{fig:FWHM} show the fit of this simple model to the experimental values of the FWHM, assuming that $\sigma$ is not affected by the deposition of neither FLG nor NiAu. From this analysis we deduce that $\sigma\approx 3.06 \pm 0.1 \times 10^{-11}$~meV$\times$cm$^{2}$ and that $\Gamma_{i}$ increases by approximately $50$~\% both under FLG and under NiAu. Thus, deposition of FLG results in a similar disorder-induced broadening of the IX transition, as compared to NiAu. 

It is evident from Fig.~\ref{fig:FWHM} that we could reach lower IX densities in NiAu and FLG- covered regions than under the bare surface. 
There are three reasons for that.
First, because these former areas are characterised by higher in-plane potential, IXs drift away from the NiAu- and FLG- covered regions towards  the bare surface regions that correspond to a lower in-plane potential, see Fig.~\ref{fig:intro}~(b).
Second,  under the bare surface IXs have the longest radiative lifetime, and thus accumulate more efficiently. 
Finally, for IXs under the bare surface  electron-hole overlap is the lowest, see Table~\ref{tab:ParamaterFit}.  One can see from Eq.~\ref{eq:I_n} that this leads to a reduction of the emission intensity for a given IX density, making measurements of the IX emission  under the bare surface  more time-consuming than under FLG or NiAu.



\subsection{IX transport in the FLG-patterned environment}
\label{subsec:transport}

\begin{figure}
	\includegraphics[width=3.4in]{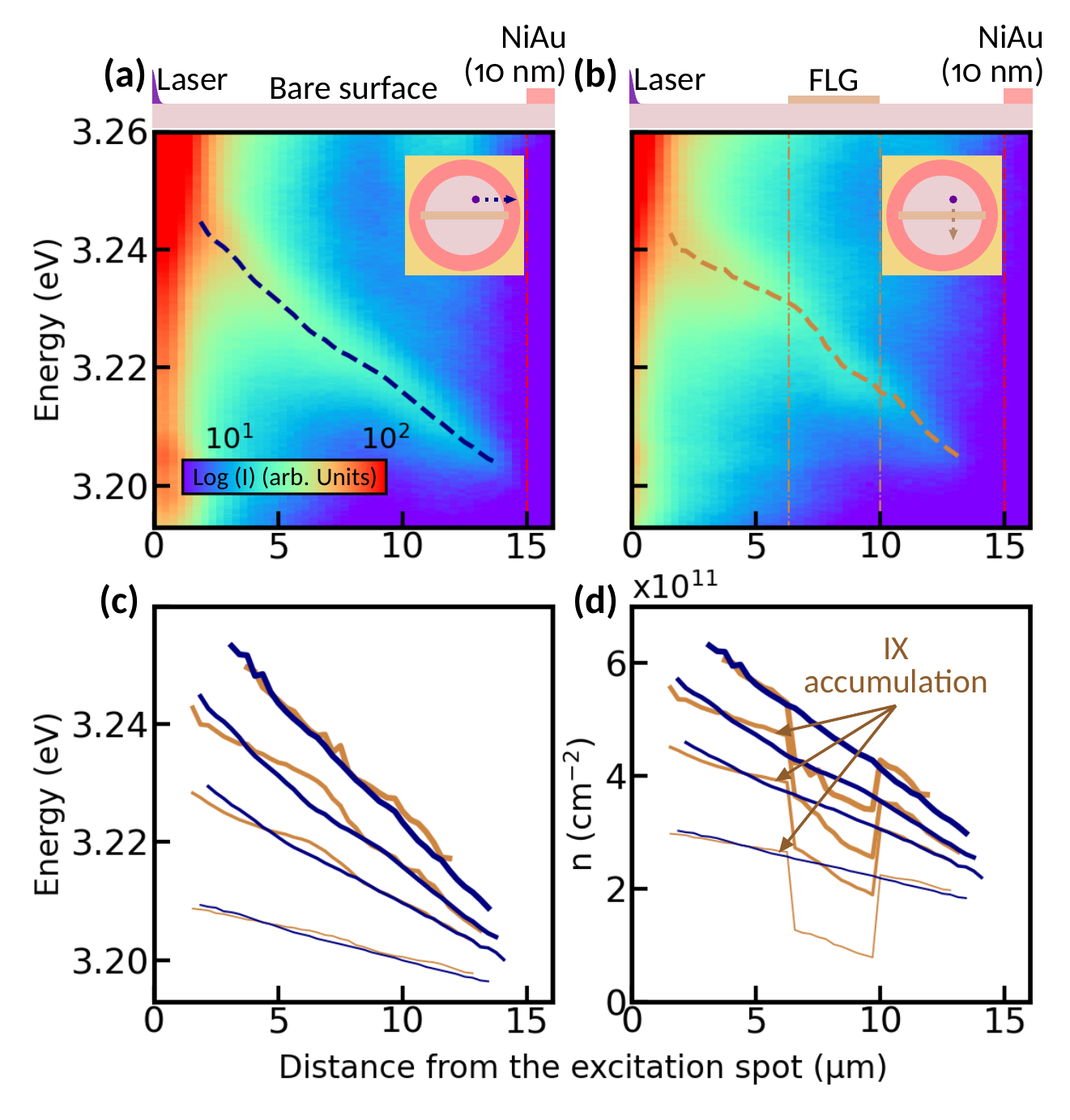}
	\caption{(a-b) PL spectra (color-encoded in log scale) measured along two perpendicular directions, either parallel (a), or orthogonal (b) to the FLG stripe (see insets). The corresponding surface state and the laser position are sketched above each spectral map. Excitation power density are $P=40$~kW/cm$^2$. Dashed lines highlight IX emission. 
 (c-d) IX energies (c) and densities (d) extracted from the data as in (a-b) at different power densities ($P = [5, 20, 40, 80]$~kW/cm$^2$.) Blue (brown) lines show IX transport in parallel (perpendicular) to FLG geometry. Arrow point the IX accumulation region in the perpendicular geometry.
	}
		\label{fig:Transport}
\end{figure}

The experiment designed to bring up the evidence of IX accumulation induced by FLG deposition is sketched in Fig.~\ref{fig:Transport}~(a-b), insets. 
The laser spot is positioned inside the circular aperture of NiAu, and IX emission at different excitation powers is resolved in two different directions: either along (a) or perpendicular to (b) the FLG stripe. 

%

Fig.~\ref{fig:Transport}~(a-b) shows color-encoded PL spectra, measured at various distances from the excitation spot (situated at $x=0$) in these two configurations, at power density $P = 40$~kW/cm$^{2}$.
Like in the case of the excitation on top of NiAu (Fig.~\ref{fig:SpectralMaps}), the emission at the excitation point is  broad and the IX-related spectrum cannot be separated from the other features. 
%
%
In contrast, at distances ranging from $2$ to $3$~$\mu$m from the spot, the emission is predominantly governed by IXs, exhibiting a characteristic decrease in energy as the distance increases.
In Fig.~\ref{fig:Transport}~(a-b) it is highlighted by dashed lines.

On can see that, when IXs travel along FLG-covered area without crossing it (a), their energy decreases smoothly as they spread away from the excitation spot. In the opposite direction (b), a gradual decline in energy, particularly before the FLG-covered region, is followed by a pronounced drop.
%
This specific IX energy profile is a footprint of the FLG-induced potential barrier, that alters IX transport.
%
Similar analysis for different excitation powers allows us to plot in Fig.~\ref{fig:Transport}~(c) the distance dependence of the IX energy in the two geometries under consideration: with (brown line) and without (blue line) FLG-induced barrier. 
The values of $E_0$ obtained in the previous section and Eq.\ref{eq:E_IX} allow us to track the IX densities in the two configurations. 
They are represented in Fig.~\ref{fig:Transport}~(d) with the same color code. It appears that the spatial profile of the IX density is, indeed, modified by the FLG. 
Two features can be identified: (i) IXs density distribution is flatter in the area preceding FLG as compared to the case of the free propagation; (ii) IX density drops down under FLG.

However, the effect is  virtually absent at the highest power density $P=80$~kW/cm$^2$.
Indeed, as discussed in Section~\ref{subsec:PotentialBarriers}, at $n>1.7\times 10^{11}$~cm$^{-2}$ IXs can completely screen out the in-plane potential barrier induced by FLG. 
Here we deal with densities around $2-6\times 10^{11}$~cm$^{-2}$. 
Under such conditions, the FLG-induced barrier cannot completely impede IX transport. IXs emit light from the FLG-covered area (although their  density is much smaller), and beyond the barrier ($x>10$~$\mu$m). 
Further studies at lower excitation powers should determine the relevant FLG geometries and IX density regime that would optimise the creation of thermalized IX fluids with different, but constant, densities.

\section{Summary and Conclusions}
\label{sec:conclusions}

We have studied the effect of FLG and NiAu deposition on the surface of the device containing a single wide GaN/(Al,Ga)N QW. 
We evaluated the resulting modifications of the surface potentials and built-in electric fields across the structure that are related to the  blue shifts of the transition energy of IXs hosted by the QW. 

It appears that FLG deposition shifts the energy by $\Delta_{\mathrm{FLG}}= 14 \pm 3$~meV, that is $\approx 6$ times less than NiAu deposition $\Delta_{\mathrm{FLG}}= 82 \pm 5$~meV.
Our results suggest that these materials are capable of modifying the in-plane potential experienced by IXs, but with different potential step heights.
We have shown that, despite smaller potential barrier induced by FLG (it can be screened out by IXs at density $n \approx 2 \times 10^{11}$~cm$^{-2}$), it can hinder IX transport and modify their density distribution in the plane of the QW. 

We also estimated the IX transition broadening, as well as the transparency losses.
While the inhomogeneous broadening increases by about a factor of two for both kinds of surface layers, this effect mainly matters at low densities, where the linewidth is dominated by this effect. 
At $n > 1.7 \times 10^{11}$~cm$^{-2}$ the collisional broadening, which depends linearly on the exciton density, start to  dominate  the inhomogeneous one. 
This collisional contribution to the transition broadening seems to be unaffected by the deposition of neither FLG or NiAu. 
Finally, the transparency losses due to FLG ($30$\%) are slightly weaker than those due to NiAu ($60$\%), consistent with previous measurements in light emitting diodes. 

In conclusion, joined implementation of NiAu and FLG may constitute a  promising tool for electrostatic modulation of IX densities.
As a next step, the exploration of various combined surface patterns can be considered. 
In particular, a comparison between FLG and NiAu-based electrostatic traps for IXs could bring further insights in our understanding of exciton-exciton interactions at low temperatures, where they are subject to strong correlations and may offer a possibility to realize collective states \cite{Chester1970,Lozovik1976,Suris2016,Glazov2021}. 



\begin{acknowledgments}
This work was supported by French National Research Agency via IXTASE (ANR-20-CE30-0032) and LABEX GANEXT projects.
\end{acknowledgments}

\bibliography{Biblio}

\end{document}